# The vital role of hole-carriers for superconductivity in pressurized black phosphorus

Jing Guo[1]*, Honghong Wang[1]*, Fabian von Rohr[2], Wei Yi[1], Yazhou Zhou[1], Zhe Wang[1], Shu Cai[1], Shan Zhang[1], Xiaodong Li[3], Yanchuan Li[3], Jing Liu[3], Ke Yang[4], Aiguo Li[4], Sheng Jiang[4], Qi Wu[1], Tao Xiang[1,5], Robert J Cava[2], Liling Sun[1,5]†

[1]*Institute of Physics and Beijing National Laboratory for Condensed Matter Physics, Chinese Academy of Sciences, Beijing 100190, China*

[2]*Department of Chemistry, Princeton University, Princeton, New Jersey 08544, USA*

[3]*Institute of High Energy Physics, Chinese Academy of Science, Beijing 100049, China*

[4]*Shanghai Synchrotron Radiation Facilities, Shanghai Institute of Applied Physics, Chinese Academy of Sciences, Shanghai 201204, China*

[5]*Collaborative Innovation Center of Quantum Matter, Beijing, 100190, China*

The influence of carrier type on superconductivity has been an important issue for understanding both conventional and unconventional superconductors [1-7]. For elements that superconduct, it is known that hole-carriers govern the superconductivity for transition and main group metals [8-10]. The role of hole-carriers in elements that are not normally conducting but can be converted to superconductors, however, remains unclear due to the lack of experimental data. Here we report the first *in-situ* high pressure Hall effect measurements on single crystal black phosphorus, measured up to ~ 50 GPa, and find a correlation between the Hall coefficient and the superconducting transition temperature ($T_C$). Our results reveal that hole-carriers play a vital role in developing superconductivity and enhancing $T_C$. Importantly, we also find a Lifshitz transition in the high-pressure cubic phase at ~17.2GPa, which uncovers the origin of a puzzling valley in the superconducting $T_C$-pressure phase diagram. These results offer insight into the role of hole-carriers in developing superconductivity in simple semiconducting solids under pressure.

Black phosphorous is an archetypical band semiconductor, regarded as an analog of graphite [11], and its monolayer variant phosphorene is similar to graphene [12-14]. Recently, the electronic topological transition found in the orthorhombic semiconducting phase at 1.2 GPa [15-18] has made black phosphorus a new focus of research in condensed matter physics and material science. Bulk black phosphorous single crystals display high charge mobility [19] and thin film black phosphorous single crystals exhibit a good field effect [20, 21], suggesting potential applications in electronic and optoelectronic devices.

Structurally, bulk black phosphorous crystallizes in an orthorhombic structure with buckled layers bonded to each other by van der Waals forces. Because pressure can shrink the volume of solids, alter atomic distances, modify chemical bonds or even induce structural phase transitions, electronic properties can often be modified significantly under pressure [22-27]. Indeed, black phosphorous presents two pressure-induced crystal structure transitions, first from an orthorhombic phase to a rhombohedral phase, and then to a cubic phase [28-33], displaying concurrently a diversity of physical properties, in particular a rhombohedral superconducting phase at ~5 GPa and a cubic superconducting phase at ~12GPa [34]. Intriguingly, black phosphorus also shows a puzzling valley in its superconducting $T_C$ versus pressure behavior in the cubic phase region of the $P$-$T_C$ phase diagram [34]. A unified understanding for the complex superconducting behavior remains lacking. In this study, we report the results of a suite of complementary high-pressure measurements, with the aim of revealing the contributions of electron- or hole-carriers to the superconductivity in the different structural phases of pressurized black phosphorus, while detailing the superconductivity-structure-pressure phase diagram of this normally semiconducting element.

**Results**

**Sample quality, pressure-induced structural phases, and superconducting phase transitions**. The quality of the sample investigated in this study was examined through synchrotron x-ray diffraction (XRD) measurements. As shown in Fig.1a, all the peaks detected from the powdered sample obtained from ground single crystals can be well indexed in the orthorhombic structure in space group *Cmca*, in good agreement with previous results [35]. Analysis of the crystals by scanning electron microscopy (SEM) demonstrates that the surfaces of the crystals are flat, and that the thickness of the samples employed lies in the 3μm -10μm range (inset of Fig.1a). The combined results obtained from our XRD and SEM experiments indicate that the single crystals used for the present study are of high quality.

We observe that black phosphorous undergoes two structural phase transitions at high pressure. At around 5 GPa the orthorhombic (O) phase partially transforms to a rhombohedral (R) phase that coexists with the O phase between 5 and 8.66GPa and then becomes a single phase at higher pressure. Finally, the R phase converts to a simple cubic (C) phase at 12.44GPa (Fig.S1, Supplementary Information).

Figures1b–1edisplay the temperature dependence of the electrical resistance at different pressures. It is seen that a semiconductor-to-metal transition occurs at 2.34GPa (Fig 1b and inset). Upon increasing the pressure to 4.93 GPa, a resistance drop is found at ~3.2K (Fig.1c) that shifts to higher temperature with increasing pressure. Zero resistance is attained at pressures higher than 9.56GPa.The onset temperature of the resistance drop exhibits a complex evolution with elevating pressure (Fig.1d and Fig.1e) and, intriguingly, exhibits a minimum at a pressure of~18.81GPa (Fig.1d and inset).We also observed this unusual evolution of the resistance-temperature-pressure behavior in an independent measurement (Fig. S3, Supplemental Information), and it has also been detected by another group [34]. The observed resistance drops are confirmed to result from superconducting transitions through our alternating-current (*ac*) susceptibility measurements (Fig.S4, Supplemental Information). The pressure-induced structural phase and superconducting phase transitions observed in our samples are consistent with earlier reports [28，32，34]. To learn more about the observed complex behavior, we carried out high pressure studies on the superconductivity of black phosphorus through Hall effect measurements.

**Pressure-temperature phase diagram**. We summarize our experimental results on black phosphorous in a pressure-temperature phase diagram (Fig.2a). This phase diagram includes information about the structural phase transitions and the superconducting transitions. The corresponding Hall effect data can be found in Fig. 2b.From the perspective of overall electronic behavior, there are three distinct regions in this phase diagram: a semiconducting phase, a semimetal phase and a superconducting phase. For pressures between 1 bar and ~5 GPa, orthorhombic black phosphorous undergoes a transition from a semiconducting state to a semimetal state at a critical pressure ($P_{C1}$) of ~2.34GPa. When the pressure is higher than ~5 GPa, the orthorhombic form partially converts to the R phase, as we described above, (Fig.S1 and Fig.S2, Supplementary Information) and superconductivity simultaneously appears, indicating that the superconductivity is induced by the O-R structural phase transition. Since the O-R transition is sluggish, the two phases coexist until 8.66 GPa. At a pressure of~12.4 GPa, black phosphorus transforms to the C phase, and the superconducting transition temperature $T_C$ shows a jump at the R-C transformation. Upon increasing pressure, $T_C$ decreases, reaching a minimum at ~ 18.81GPa in the C phase, and then increases again with further elevating pressure, forming a superconducting valley.

**The evolution of Hall resistance and Hall coefficient with pressure**. To understand the pressure-induced changes of the superconducting transition temperature in black phosphorus, we performed high-pressure Hall resistance measurements by sweeping the magnetic field (*B*) from 0T to 7Tperpendicular to the *ab* plane of a single crystal sample at 15 K (close to the superconducting transition temperatures) as shown in Fig. 3. We also derived the Hall coefficient ($R_H$) from the Hall resistance $R_{xy}$ for each pressure point and established the pressure dependence of $R_H$ (Fig.2b). It is noted that the sign of $R_{xy}$ is the same as that of $R_H$ (Supplementary information); thus the correlation between superconductivity and dominant electron- or hole-carriers can be established. In the orthorhombic semiconducting phase, the value of Hall resistance $R_{xy}$ (or $R_H$) is positive, reflecting the dominance of hole-carriers in this pressure range, but $R_{xy}$ becomes negative below a magnetic field of 3.55T at 2.34GPa (Fig.3b). On further pressurization, we find that $R_{xy}(B)$ completely reverses to a negative slope in the range between 2.94 GPa and 7.81 GPa (Fig 3b), and, unexpectedly, that it changes its slope again, this time from negative to positive, in the pressure range of 9.56 GPa-10.9 GPa, where black phosphorus fully enters the R phase (Fig.3c, and Fig.S1 of Supplementary Information). At the pressure of 12.46 GPa, where black P converts to the C phase, $R_{xy}(B)$ or $R_H$ is negative (Fig.3c and Fig.2b). At 16.95 GPa, $R_{xy}$ is almost zero below 2T, above which the slope of $R_{xy}(B)$ increases further upon elevating field up to 23.37 GPa and then decreases with further increasing pressure (as indicated by arrows in Fig.3d). This complicated variation of Hall resistance and Hall coefficient with pressure reveals that the topology of the Fermi surface of black phosphorous undergoes substantial changes in the pressure range investigated. This is the first report of Hall measurements for the superconducting phases of black phosphorous.

**Discussion**

Next we focus on the analysis of the correlation between superconductivity and Hall coefficient in the R phase and C phase.

**Superconductivity and Hall coefficient in the rhombohedral phase.** In the mixed phase (O+R phase) region, the $R_{xy}(B)$ of black phosphorus displays non-linearity (Fig.3b), though its $R_H$ is negative, implying that there exist two types of carriers (electron and hole) at the Fermi surface. The negative $R_H$ observed in the range of the mixed phase demonstrates that the electron-carriers are dominant. However, $R_H$ becomes positive when black phosphorus transforms to a single R phase (Fig.2b). More importantly, $R_{xy}(B)$ of the R phase presents linear behavior for magnetic fields below 7T (Fig.3c). These results suggest that only one type of hole-carrier is found in the R phase [36], implying that these hole-carriers play a vital role in developing superconductivity in the R phase. Since the absolute value of the negative $R_H$ in the mixed phase is much higher than that of the positive $R_H$ in the single R phase (Fig.2b), it can be concluded that the electron-carriers detected in the mixed phase region are mainly contributed by the non-superconducting orthorhombic phase. We find that the superconducting $T_C$ in the R phase increases as the population of the hole-carriers increases, indicating that higher concentrations of hole-carriers favor higher superconducting $T_C$s in the R phase.

**Lifshitz transition and its correlation with the superconducting valley in the cubic phase.** At pressures of ~12 GPa, black phosphorus converts from the R phase to the C phase. The superconducting $T_C$ shows a jump at this pressure (Fig.2a). Simultaneously, $R_H$ changes its sign from positive to the negative again (Fig.2b). Our $R_{xy}$ measurements demonstrate that the linear slope of $R_{xy}(B)$ is broken at $P > 12$GPa (Fig.3c), and that a non-linear $R_{xy}$ with respect to $B$ takes place in the C phase (12-50 GPa), revealing the coexistence of electron and hole carriers. Significantly, within the single C phase, the sign of $R_H(P)$ changes from negative to positive at $P_{C2}$ (~17 GPa), near the pressure where the superconducting valley is observed (Fig.2a and 2b). This sign change of $R_H$ in the absence of a structural phase transition can be characterized as a Lifshitz transition, reflecting the reconstruction of the Fermi surface due to a relative population change between hole and electron carriers [15,17,37-38]. A previous study of powdered black phosphorus also found such a superconducting valley in the cubic phase [34], but the physics behind that observation has not previously been clear. Our study reveals that the origin of this valley is a Lifshitz transition. A superconducting transition associated with a Lifshitz transition has been found in many compounds [37,38], but, to our knowledge, this is the first time it has been observed in a solid element. Electronic structure calculations on cubic black phosphorus indicate that there are both large and small Fermi surfaces near the Fermi energy [39-41]; the balance between these may easily change under pressure.

In summary, our temperature-pressure phase diagram together with the evolution of structure and Hall coefficient provides insight into the importance of hole-carriers in developing superconductivity and enhancing $T_C$ in pressurized black phosphorus. Our results demonstrate that once hole-carriers become dominant in the rhombohedral and cubic phases, the superconducting $T_C$ of black phosphorus is enhanced. A Lifshitz transition, in the cubic superconducting phase, is the first of its kind in a superconducting element; we interpret the puzzling superconducting valley as being due to this Lifshitz transition. These results provide new fundamental information for shedding light on the physics of superconductivity in non-metallic elements that are transformed into metals.

**Methods**

**Single crystal growth.** Large, high-quality crystals of black phosphorus were obtained by a vapor transport technique [42]. Dry red phosphorus (150 mg, purity 99.99%), dry $SnI_4$ (10 mg, purity 99.999%), tin shot (40 mg, purity 99.8 %), and gold shot (20 mg, purity 99.999 %) were sealed in an evacuated quartz glass tube. The mixture was heated to 680 °C and held at this temperature for 10 h. It was then cooled with 0.1 K/h to 500 °C, and quenched to room temperature.

**Experimental details**

Pressure was generated by a diamond anvil cell with two opposing anvils sitting on Be-Cu supporting plates. Diamond anvils with 300 μm flats and non-magnetic rhenium gaskets with 100 μm diameter holes were employed for different runs of the high-pressure studies. The standard four-probe method was applied on the *ab* plane of single crystal black phosphorus for the high-pressure resistance measurements, and the Van der Pauw method was used for high-pressure Hall measurements. To keep the sample in a quasi-hydrostatic pressure environment, NaCl powder was employed as a pressure transmitting medium for the resistance and Hall coefficient measurements. High-pressure alternating-current (*ac*) susceptibility measurements were conducted using home-made primary/secondary-compensated coils that were wound around a diamond anvil [43,44].

Ambient pressure and high pressure X-ray diffraction (XRD) measurements were carried out at beamline 4W2 at the Beijing Synchrotron Radiation Facility and at beamline 15U at the Shanghai Synchrotron Radiation Facility, respectively. Diamonds with low birefringence were selected for the X-ray experiments. A monochromatic X-ray beam with a wavelength of 0.6199 Å was employed. To maintain the sample in a hydrostatic pressure environment, silicon oil was used as a pressure transmitting medium. Pressure was determined by the ruby fluorescence method [45].

**Acknowledgements**

The authors would like to thank Prof. Zhongxian Zhao for stimulating discussions. The work in China was supported by the NSF of China (Grants No. 91321207, No. 11427805, No. 11404384, No. U1532267, No. 11604376), the Strategic Priority Research Program (B) of the Chinese Academy of Sciences (Grant No.

XDB07020300) and the National Key Research and Development Program of China (Grant No.2016YFA0300300). The work at Princeton was supported by the Gordon and Betty Moore Foundation EPiQS initiative, grant GBMF-4412.

†Corresponding authors

llsun@iphy.ac.cn

* These authors are contributed equally.

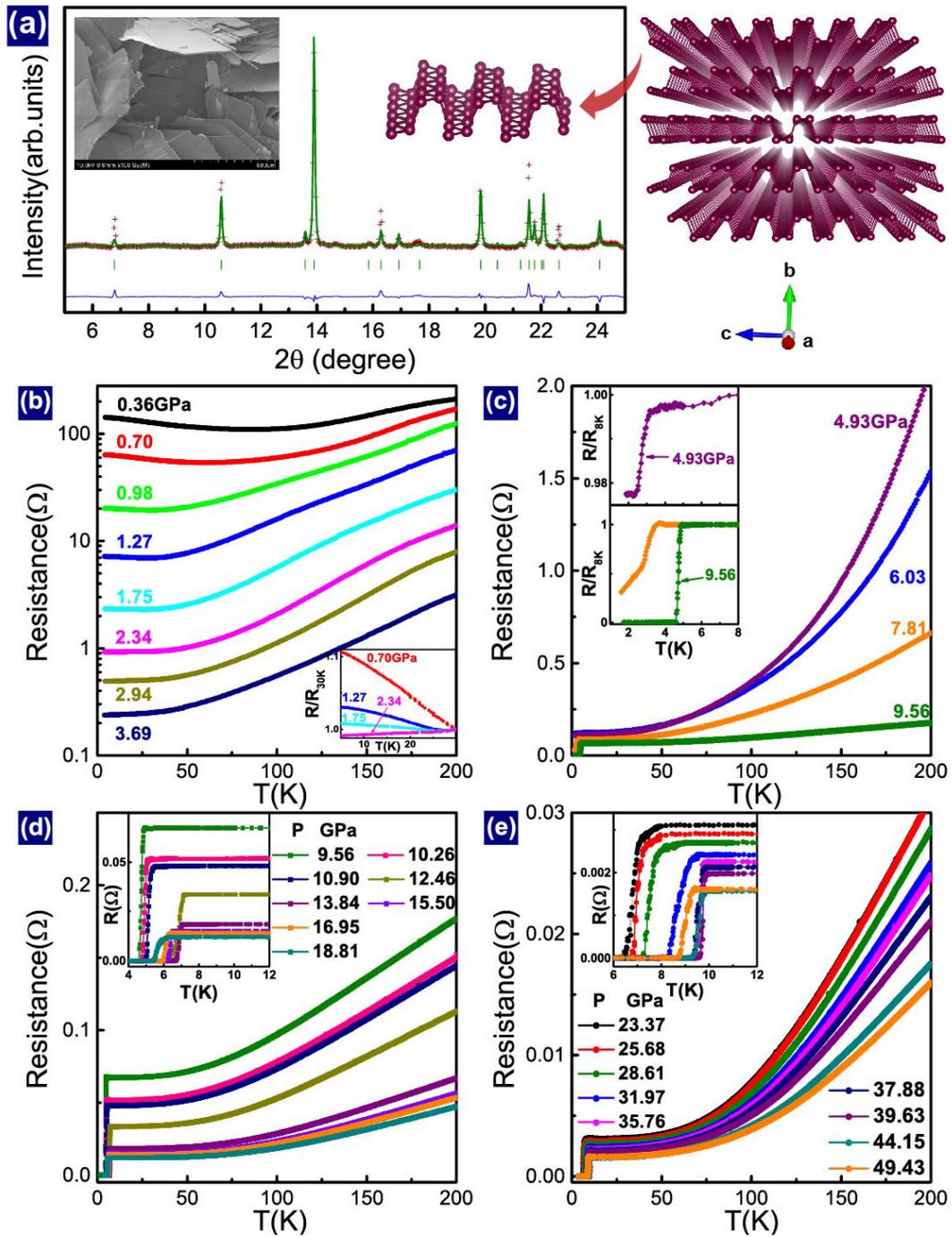

**Figure 1 Structural information for black phosphorus and its electrical resistance at different pressures.** (a) Experimental, indexed X-ray diffraction patterns of black phosphorus at ambient pressure. The red crosses represent experimental data, and the green line and bars represent the calculated peak positions for the orthorhombic phase. The left inset is a scanning electron microscope (SEM)

image of the as-grown black phosphorus single crystals. The right upper figure shows the crystal structure of orthorhombic black phosphorus. (b) - (d) Electrical resistance as a function of temperature for single crystal black phosphorus at different pressures. The insets display enlarged views of the resistance or normalized resistance in the lower temperature range around the superconducting transition.

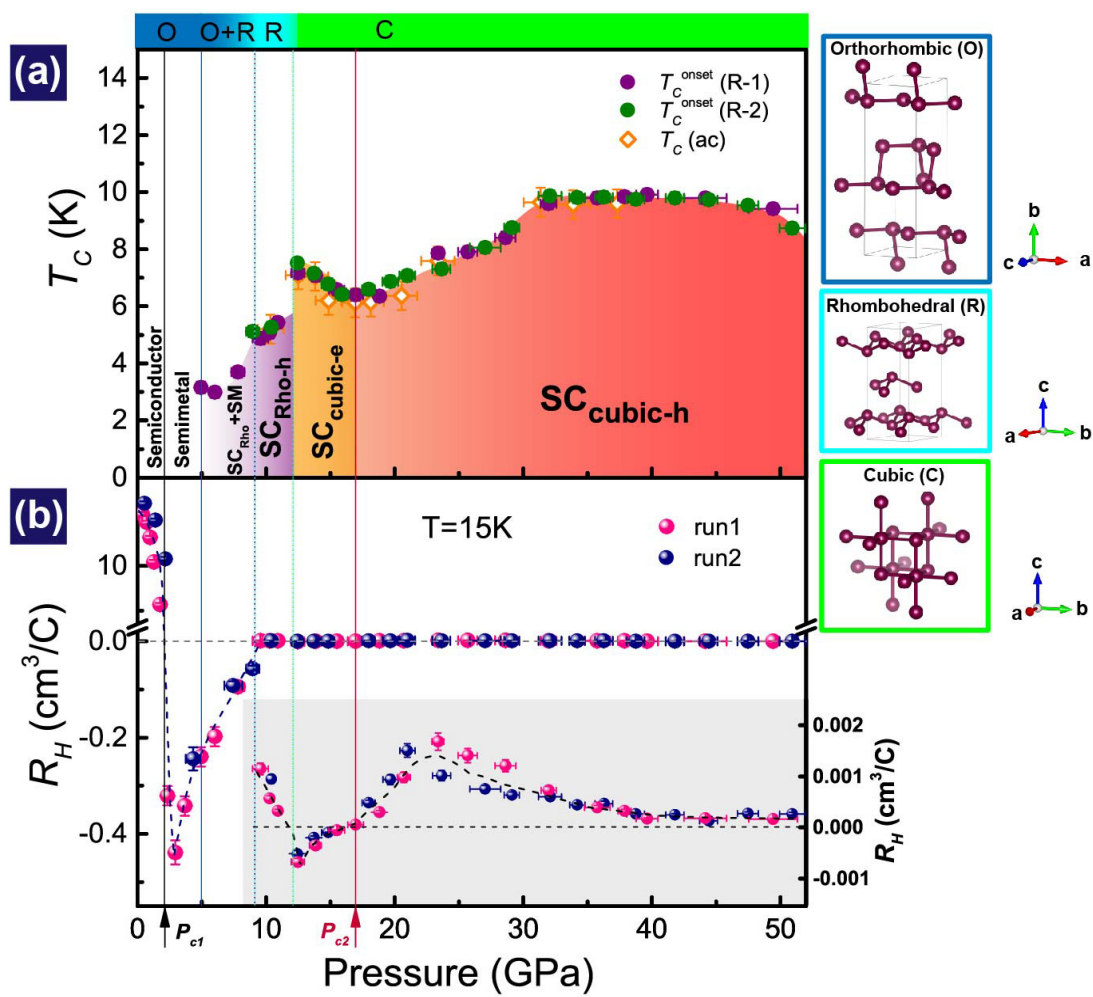

**Figure 2 Structural, superconducting and Hall coefficient ($R_H$) phase diagram of black phosphorus at different pressures.** (a) Pressure-Temperature phase diagram combined with structural phase information. Here SM and $SC_{Rho-h}$ represent semimetal and superconducting phases with rhombohedral structure and dominant

hole-carriers, respectively. SC$_{cubic-e}$ and SC$_{cubic-h}$ stand for superconducting phase with cubic structure and dominance of electron-carriers and hole-carriers, respectively. $T_C^{onset}$(R-1) and $T_C^{onset}$(R-2) represent the onset temperature of superconducting transitions determined from resistance measurements for run-1 and run-2. $T_C(ac)$ represents the superconducting transition temperature determined by *ac* susceptibility measurements. (b) Pressure dependent $R_H$ measured at 15K. The inset displays the detailed changes of $R_H$ at the structural phase transition boundaries. The pink and blue solid points represent Hall coefficients obtained from two independent runs. $P_{c1}$ and $P_{c2}$ stand for the critical pressures of the Lifshitz transitions.

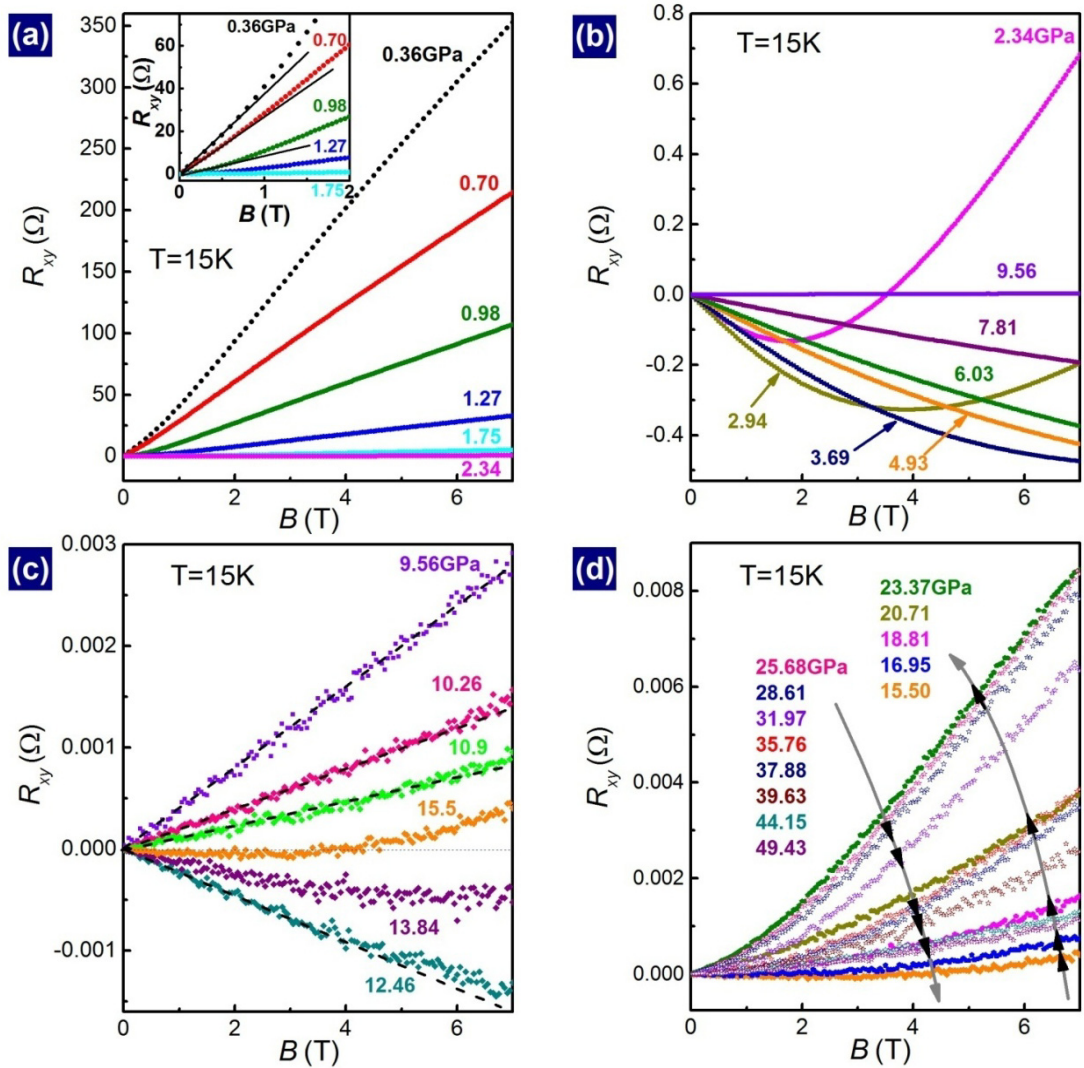

**Figure 3 Hall resistance ($R_{xy}$) versus magnetic field ($B$) at 15 K and different pressures for black phosphorus single crystals.** Plots of $R_H$-$B$ in the pressure range of (a) 0.36GPa-2.34 GPa, (b) 2.34 GPa - 9.56 GPa, (c) 9.56 GPa -15.5 GPa and (d) 15.5 GPa - 49.43 GPa.